\documentclass[epj]{svjour}

\usepackage{graphicx,amssymb,amsmath}
\usepackage[applemac]{inputenc}

\begin{document}

\title{Wave turbulence buildup in a vibrating plate}
\author{Maria Ines Auliel\inst{1,2}, Benjamin Miquel\inst{3}, and Nicolas Mordant\inst{2,4}}
\institute{Laboratorio de Fluidodinamica, CONICET- Facultad de Ingenieria, UBA, Bs. As,
Argentina and Universidad Nacional de Tres de Febrero, Caseros, Prov. de Bs. As, Argentina. \and
Laboratoire des Ecoulements G\'eophysiques et Industriels, Universit\'e Grenoble Alpes, CNRS  F-38000 Grenoble, France. \and
Department of Applied Mathematics, University of Colorado, Boulder, CO 80309-0526, USA.\and
Institut Universitaire de France, 103, bd Saint Michel, F-75005 Paris, France.}



\abstract{We report experimental and numerical results on the buildup of the energy spectrum in wave turbulence of a vibrating thin elastic plate. Three steps are observed: first a short linear stage, then the turbulent spectrum is constructed by the propagation of a front in wave number space and finally a long time saturation due to the action of dissipation. The propagation of a front at the second step is compatible with scaling predictions from the Weak Turbulence Theory.}

\maketitle

Wave turbulence is a generic state involving a large number of non linearly interacting waves. The typical example is that of the spectrum of sea surface waves which involves a wide interval of wavelengths and thus of frequencies. In the limit of vanishing wave amplitudes and infinite system size, a statistical theory can be developed that predicts the evolution of the wave spectrum: the Weak Turbulence Theory (WTT)~\cite{R1,R2,R3}. For small amplitudes, the nonlinear transfer of energy among waves is very slow as compared to the wave frequency. Thanks to this scale separation, a multi scale analysis can be developed that leads to the so called ``kinetic equation''. This equation describes the slow temporal evolution of the wave spectrum. Formally, energy is transferred through the ``collision'' of at least 3 resonant waves. A major advance was made by Zakharov by exhibiting stationary solutions of the wave spectrum in the out of equilibrium case, i.e. in presence of forcing and dissipation. In the ideal case of forcing localized at large scale and dissipation operating at small scales, the solutions show that a range of wavelengths can exist in which energy cascades conservatively in wavenumber space. This 
energy cascade is similar to that of fluid turbulence~\cite{Frisch}. In the latter case no statistical theory supports the famous $k^{-5/3}$ Kolmogorov spectrum. For wave turbulence, the stationary wave spectra can be calculated from the theory and are called Kolmogorov-Zakharov (KZ) spectra.

The predictions for the stationary KZ spectrum have been calculated for many physical systems such as magnetized plasmas (solar winds or tokamaks), sea surface (gravity and capillary waves), light in nonlinear media, Kelvin waves on superfluid quantum vortices... (see \cite{R1,R2,R3} for reviews). By contrast, laboratory observations are scarce and concern mostly water waves~\cite{R10,R11} but also inertial waves~\cite{Sharon} or light~\cite{R8}. Recently an experimental breakthrough occurred concerning the case of flexural waves in a thin elastic plate~\cite{Boudaoud,Mordant,R19,R21,R23,R24,R25,Miquel3,Humbert}. In this system, it is possible to implement a space and time resolved measurement of the wave field that allows us to probe the theory in depth~\cite{R19}. The WTT has been applied to this case by D\"uring {\it et al}~\cite{R18}. The equation of motion of a thin elastic plate are the F\"oppl-von K\'arm\'an equations:
\begin{eqnarray}
\partial _{tt} \zeta &=& -\frac{E h^2}{12\rho(1-\sigma^2)}\Delta^2\zeta + \frac{1}{\rho}\left\{\zeta,\chi \right\}\\
\Delta^2 \chi &=&- \frac{E}{2\rho}\left\{\zeta, \zeta \right\} \,\, ,\label{eq_FVK}
\end{eqnarray}
where the physical properties of the material are described by the following coefficients: Young's modulus $E$, Poisson's ratio $\sigma$, the density $\rho$. The brackets $\left\{\cdot,\cdot\right\}$ denote the bilinear differential operator 
\begin{equation}
\left\{\zeta,\chi\right\}=\partial_{xx}\zeta\partial_{yy}\chi + \partial_{yy}\zeta\partial_{xx}\chi -2 \partial_{xy}\zeta\partial_{xy} \chi\,.
\end{equation}
The linear part of the wave equation provides the dispersion relation for vanishingly small wave amplitudes that are only due to flexion: $\omega=ck^2$ with $c=\sqrt{\frac{E h^2}{12 \rho(1-\sigma^2)}}$. The non linearity is due to in plane stretching of the plate at large deformation.

The WTT formalism has been applied to these equations and leads to the following kinetic equation~\cite{R18}:
\begin{eqnarray}
\frac{dn_{\mathbf k}}{dt}&=&\int 
\sum_{s_1s_2s_3}n_{\mathbf k_1}n_{\mathbf k_2}n_{\mathbf k_3}n_{\mathbf k}
\left (\frac{1}{n_{\mathbf k}}+\frac{s_1}{n_{\mathbf k_1}}+\frac{s_2}{n_{\mathbf k_2}}+\frac{s_3}{n_{\mathbf k_3}}\right)\nonumber\\
&&\times\delta(\omega_{\mathbf k}+s_1\omega_{\mathbf k_1}+s_2\omega_{\mathbf k_2}+s_3\omega_{\mathbf k_3})\nonumber\\
&&\times\delta(\mathbf k+s_1\mathbf k_1+s_2\mathbf k_2+s_3\mathbf k_3)\nonumber\\
&&\times12\pi |J_{\mathbf k,\mathbf k_1,\mathbf k_2,\mathbf k_3}|^2d^2\mathbf k_1d^2\mathbf k_2d^2\mathbf k_3 \,\, ,
\label{ke}
\end{eqnarray}
where $n_{\mathbf k}(t)$ is the wave action at wave vector $\mathbf k$ (such that $\omega_{\mathbf k}n_{\mathbf k}$ is the wave energy spectrum). $J$ is an interaction kernel which expression can be found in~\cite{R18}. Stationary spectra cancel the collision integral on the right-hand side of the kinetic equation. Previous work have shown that the wave spectra observed in real vibrating plates differs from the theoretical prediction of the KZ spectrum:
\begin{equation}
n^{KZ}(\mathbf{k}) \propto \frac{\ln^{1/3} {(k^*/k)}}{k^2}\,\, ,
\end{equation}
with $k=||\mathbf k||$ and $k^*$ is a cutoff wavenumber. A careful analysis of the experimental wave motion was possible due to the space and time resolved measurement~\cite{R19,R21,R23,R24,R25,Miquel3}. It showed that the observed wave turbulence was truly weakly non linear turbulence but that the discrepancy with the theory was due to the dissipation. Indeed in real plates dissipation is not ideally localized at high wave numbers as assumed in the theory but it is rather distributed at all scales~\cite{R23,Humbert}. Thus the theory has to be adapted to take into account the existence of such realistic dissipation. Wave turbulence can still develop if dissipation is so small that a double timescale separation exists between the fast linear oscillations, intermediate non linear time scales and very slow dissipation timescales~\cite{Miquel3}. 

Although the stationary case seems to be consistent with the theory of elastic plates, at least in the phenomenology, the transient buildup of the spectra has almost never been investigated experimentally. The few results concern non linear optics~\cite{R8} or water waves~\cite{Bedard}. Predictions have been made that the buildup of the wave spectrum should follow a self similar evolution in which a front propagates in spectral space from large to small scales~\cite{Falkovich}. A first numerical result has been reported by Ducceschi {\it et al.} for non dissipative finite plates that confirms that such a front propagates~\cite{Ducceschi}. We propose to investigate this transient evolution experimentally and numerically in realistic plates. We report first experimental results of the buildup of the wave spectrum when starting the forcing in a real plate (part I). Then we complement these observations by numerical simulations (part II) used to check the robustness of the experimental observations in a idealized setup.

\section{Experiments}

\subsection{Description of the experimental setup}

\begin{figure}[!htb]
\centering
\includegraphics[width=7cm]{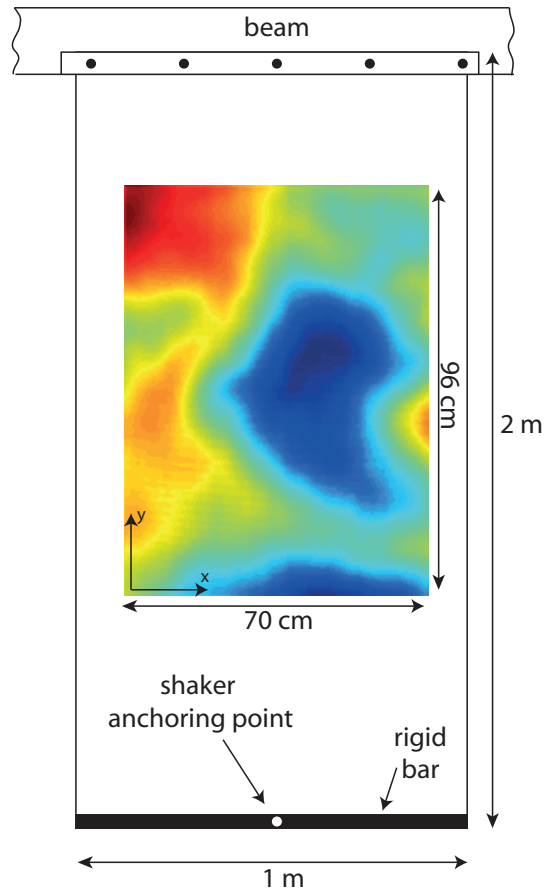}
\caption{Schematics of the experiment. A $2\times1$ m$^2$ stainless steel plates is hung vertically on a beam. Vibration is generated by an electromagnetic shaker anchored on a rigid bar fixed on the bottom edge of the plate. The deformation of the plate is measured in the center region by a Fourier transform profilometry technique.}
\label{setup}
\end{figure}

The experimental setup is close to that of~\cite{R21}. A stainless steel plate hangs vertically under its own weight (fig.~\ref{setup}). Its dimensions are: 2$\times$1~m$^2$ and 0.5~mm thick. The plate used in the previous references was 0.4~mm thick (with the same size). This change alters slightly the dispersion relation of the waves but does not change significantly the observed turbulence. The plate is clamped at the top along its short edge. A more important change concerns the connection between the plate and the electromagnetic shaker: here the shaker is attached at the very bottom of the plate and the edge of the plate is rigidified by an aluminum L bar: thus the shaker imposes a uniform normal velocity over the entire edge. It induces a vertically propagating plane wave at 30 Hz (same excitation frequency as in the previously reported experiments) that corresponds to a wavenumber $k/2\pi=2.6$~m$^{-1}$. This change of the forcing geometry was implemented in order to have mostly unidirectional forcing scheme. It makes the initial stage of the transient regime easier to understand as will be shown below.

The deformation of the plate is measured both in space and time using a Fourier transform profilometry technique as described in~\cite{R22,Cobelli1}. The deformation is measured over a surface $70\times 96$~cm$^2$. The principle of the measurement is the following: a gray scale pattern is projected on the plate (painted mate white) by a videoprojector. This pattern is then visualized by a high speed camera. When the surface is deformed, the pattern recorded by the camera is deformed as well. The chosen specific pattern is a grayscale sine modulation. The deformation of the plate can be calculated to induce a phase modulation of the pattern that can be related through geometrical optics to the shape of the plate. The distorded pattern is demodulated by means of a Hilbert transform to recover the phase which is then inverted to get the deformation of the plate. This demodulation takes advantage of FFT-based algorithms and hence can be used to process hundred of thousand of images in a reasonable timespan.

The sequence of data acquisition is the following: the camera recording is triggered at $t=-0.1$~s. Then the shaker is started at its chosen magnitude of vibration at $t=0$~s. The total recording time is 1~s at 8000 frames/s. Then the excitation is stopped and the motion damps naturally during the transfer time of the acquired images to the workstation (about 5 minutes). The previous sequence is then repeated 100 times in order to obtain independent realizations of the process. Indeed the process being non stationary in time, only averages over realizations are meaningful. 

At a given wavevector $\mathbf k$, two waves can propagate with negative and positive frequencies (resp.), i.e. propagating in the direction of the vector $\mathbf k$ or the opposite direction respectively. Thus when computing the Fourier transform $\zeta^H(\mathbf k,t)$, we obtain the amplitude of the superposition of two counter-propagating waves. We separate these two waves by using the Hilbert transform \emph{in time} of the deformation of plate $\zeta(x,y,t)$~\cite{R21}. We can obtain in this way an analytic signal $\zeta^H(x,y,t)$ which spectrum at positive frequencies ($\omega>0$) is zero. In this way the Fourier transform \emph{in space} $\zeta^H(\mathbf k,t)$ at a given wave vector $\mathbf k$ corresponds to the amplitude of the wave propagating in the direction of $\mathbf k$ and not to the superposition of two counter-propagating waves. We can then study the isotropy of the wave field in more details. In the following we omit the $H$ superscript for simplicity as we use only $\zeta^H(\mathbf k,t)$ or the equivalent for the normal velocity $v=\frac{\partial \zeta}{\partial t}$.

\subsection{Buildup of the experimental spectrum}

\subsubsection{Sequence of the spectrum buildup}

\begin{figure}[!htb]
\centering
\includegraphics[width=9.5cm]{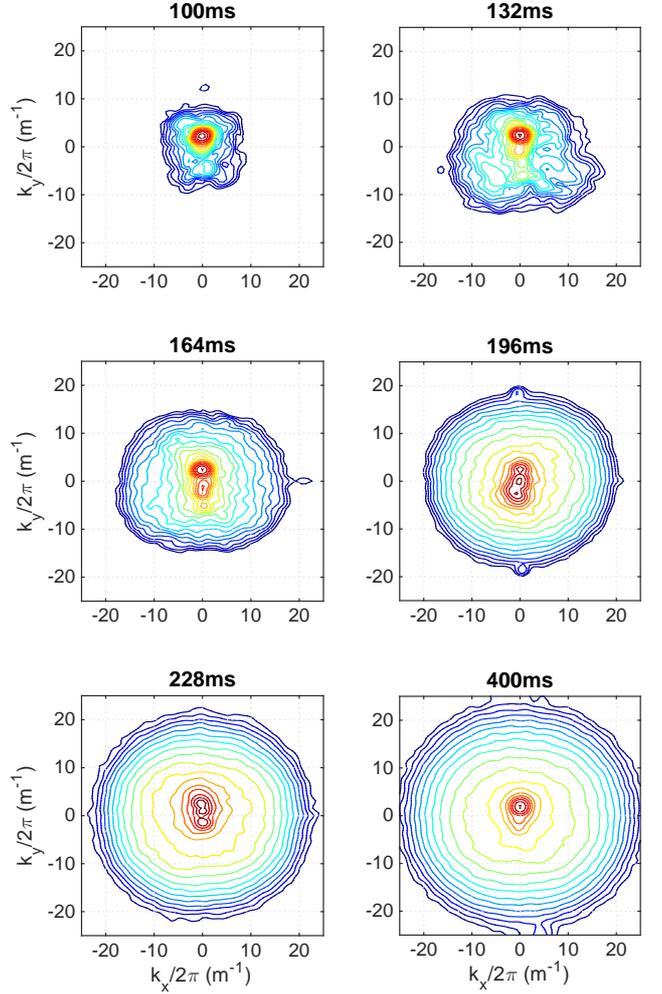}
\caption{Temporal evolution of the power spectrum of the normal velocity of the plate. The contours are spaced logarithmically between $10^{-7}$ (black) and $10^{-2}$ m$^4$/s$^2$ (red). The forcing starts at $t=0$~ms. At most times, the maximum of the spectrum corresponds to the wave number $k/2\pi=2.6$~m$^{-1}$ of the forcing wave at $30$~Hz. $t=400$~ms (bottom-right) corresponds to the stationary regime.}
\label{spec1}
\end{figure}

\begin{figure*}[!htb]
%
(a)\includegraphics[width=16cm]{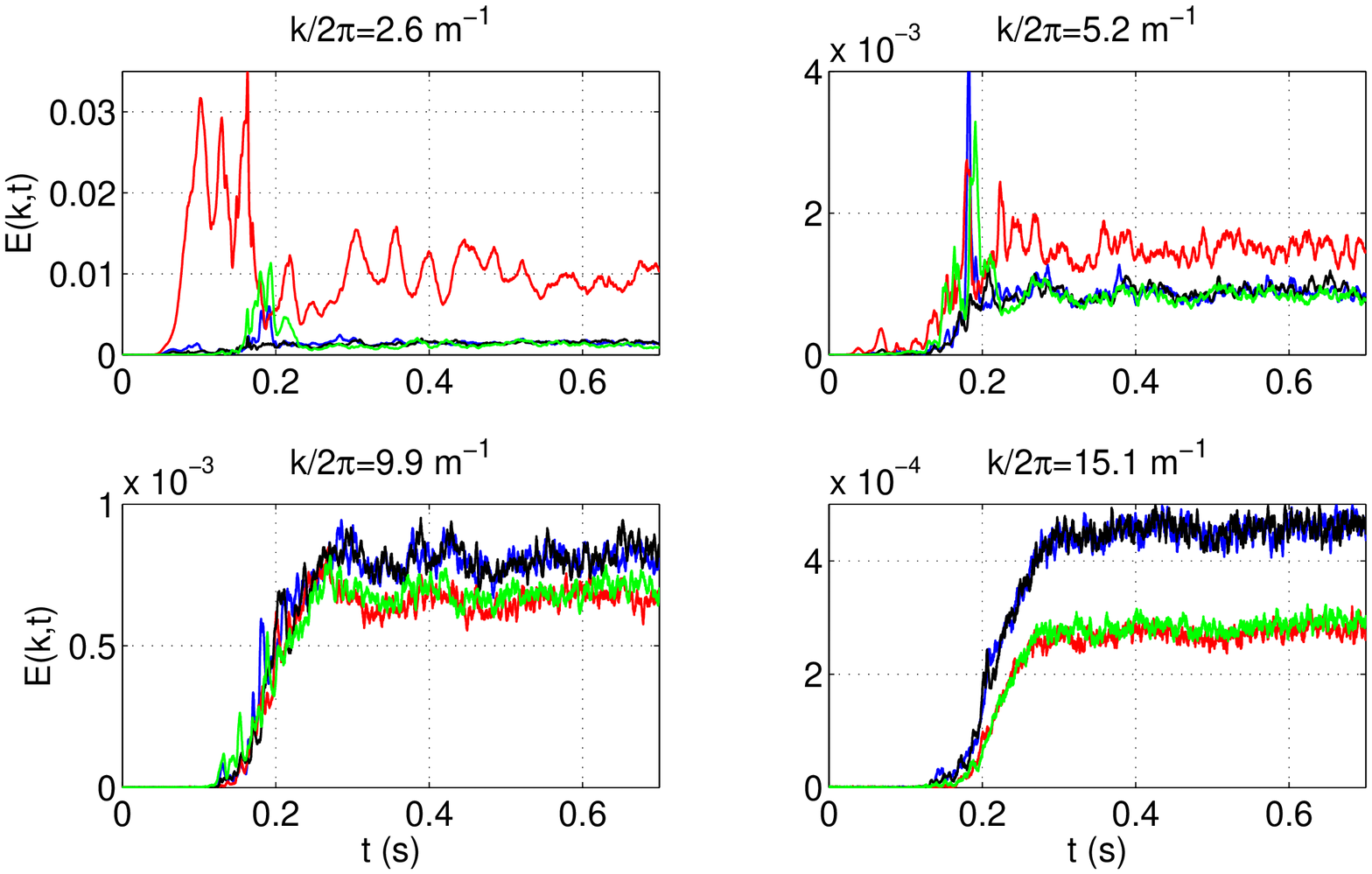}

(b)\includegraphics[width=16cm]{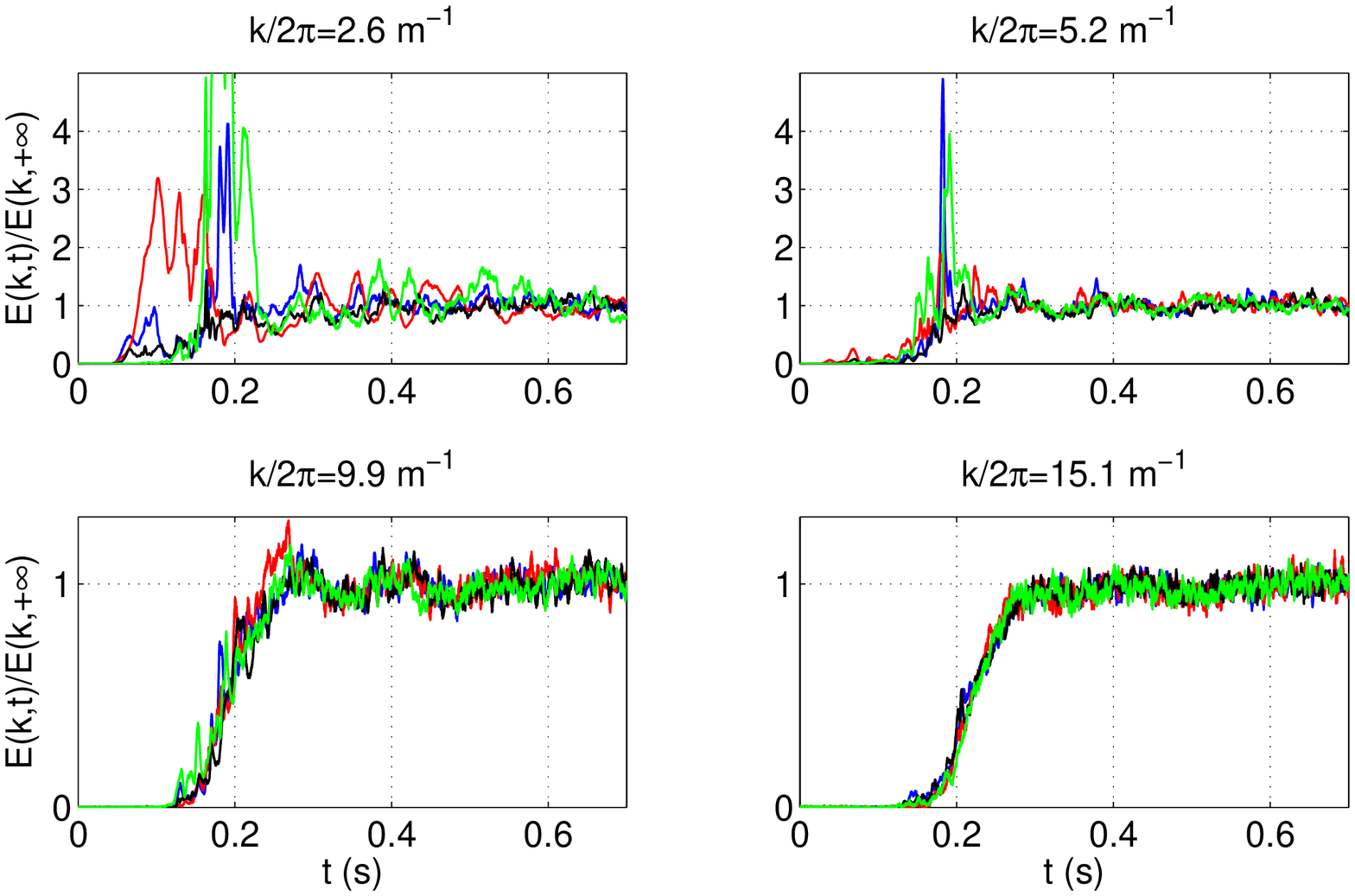}
\caption{Time evolution of the energy at various wave numbers for the 4 quadrants of propagation (blue: positive $x$ direction, red: positive $y$ direction, black: negative $x$ directions and green: negative $y$ direction). (a) selection of 4 wave numbers as indicated on each subfigure. (b) same selection but the spectrum has been normalized by the average value of the stationary limit. $k/2\pi=2.6$ m$^{-1}$ corresponds to the forced wave number.}
\label{spec2}
\end{figure*}

Figure~\ref{spec1} and \ref{spec2} display the general evolution of the spectral content of the deformation of the plate. 

At early times ($t<250$~ms) the deformation is mostly due to an upward propagating wave corresponding to the forcing frequency. This is very clearly seen in the top left part of figure~\ref{spec1} were a peak is observed for $k_y/2\pi\approx 2.6$~m$^{-1}$ and $k_x/2\pi\approx 0$~m$^{-1}$. Thanks to the Hilbert decomposition the upward propagating wave is clearly separated from the downward propagating one. 

At later times a continuous spectrum develops progressively in a quite isotropic way with a propagating front in wave number until it reaches a stationary shape (bottom-right of figure~\ref{spec1}). In the stationary regime, energy is distributed along all directions. A peak is still observed at the forcing wave number due to the fact that the forcing operates at constant frequency and amplitude and thus a temporally coherent component remains at this frequency. The spectrum is not truly isotropic at large wave numbers. Indeed the energy spectral support is slightly wider in the $x$ direction than in the $y$ direction. This is probably due to the distinct boundary conditions in these directions: the left and right boundary are free and thus the reflection coefficient of the waves is very close to one. The top boundary is clamped and the bottom one is clamped and connected to the electromagnetic shaker. These boundary conditions are most likely more dissipative because of energy leak to the support, notably to the connection to the shaker. Thus the dissipation of waves during the reflection is probably larger for waves reaching the top and bottom. This must induce the observed slight anisotropy that is also observed in fig.~\ref{spec2}.

In order to get more insight into the development of the wave spectrum we divide the spectrum in quadrants and perform an angular averaging in each quadrant. In this way the total Fourier spectrum  
$E(\mathbf k,t)=|\langle v(\mathbf k,t) \rangle|^2$ is decomposed into 4 components corresponding to polar angles in the intervals $[-\pi/4,\pi/4]$, $[\pi/4,3\pi/4]$, $[3\pi/4,5\pi/4]$, $[-3\pi/4,-\pi/4]$ and thus to directions globally oriented to the right (positive $x$ direction), to the top (positive $y$ direction), to the left (negative $x$ directions) and to the bottom (negative $y$ direction) respectively.

Figure~\ref{spec2} shows the evolution of the energy of the four quadrants for a few wave numbers ($k/2\pi=2.6$~m$^{-1}$ corresponds to the wave number at the frequency of excitation). As observed on the previous figure, the first stage of the observed motion is the propagation to the top of the plate of a planar wave directly excited by the shaker. A sharp rise of the energy of the upward motion starts at $t\approx 50$~ms (red curve at $k/2\pi=2.6$~m$^{-1}$ in fig.~\ref{spec2}). This slight delay after starting the forcing is due to the time required for the forcing wave to propagate from the bottom of the plate to the field of view of the camera. The energy of the upward forcing wave overshoots strongly the long time limit and then slowly decays to the limit. A subsequent sharp rise of the reflected downward wave starts at time $t\approx 120$~ms and peaks at $t\approx 200$~ms (green curve at $k/2\pi=2.6$~m$^{-1}$). The delay between the top and down components corresponds to the propagation over about 1.7~m at the group velocity at 30~Hz (about 24~m.s$^{-1}$) which corresponds to the propagation up to the top of the plate, bouncing and propagation downwards to the field of view of the camera. The forcing induces a noticeable anisotropy for the wavenumber $k/2\pi = 2.6$: upward propagating waves have the largest amplitude. Note that planar waves in an infinite plate cancel the non linear term of the F\"oppl-von Karman equations and thus a planar sine wave is a solution of the motion at all amplitudes. This is most likely the case until the forcing wave hits the top boundary. The upward wave, the downward reflected wave as well as the waves scattered by the boundary can then interact nonlinearly to initiate the energy cascade to smaller scales. 

At $k/2\pi=5.2$~m$^{-1}$ the energy of the modes propagating upwards remains slightly dominant. This is potentially a residual of the forcing anisotropy that would subsist in the first steps of the energy cascade. Another technical explanation could be related to the finite resolution in Fourier space that causes the peak at $k/2\pi=2.6$~m$^{-1}$ to be spread over neighboring wave vectors. The peak is significantly larger than the background and thus it influences slightly the upward wave numbers at $k/2\pi=5.2$~m$^{-1}$. For larger wave numbers the energy is the same between left and right or top and bottom propagating waves. The previously mentioned slight anisotropy between vertical or horizontal wave vectors is recovered in this figure at the highest wavenumbers.

Figure~\ref{spec2}(b) shows that once normalized by their long time limit, the energy at a given wave number (but the forcing one) increases at the same time whatever the direction of propagation. In terms of time scales, the system seems isotropic. The rise of the energy at the larger wave numbers occurs only after the rise of the downward propagating forcing wave. At the highest wave numbers, the delay of the energy rise seems to increase with the wave number in a way that is consistent with the picture of a energy front propagating  in wave number space.

\subsubsection{Deterministic part \emph{vs} stochastic part of the spectrum}

\begin{figure*}[!htb]
\centering
\includegraphics[width=16cm]{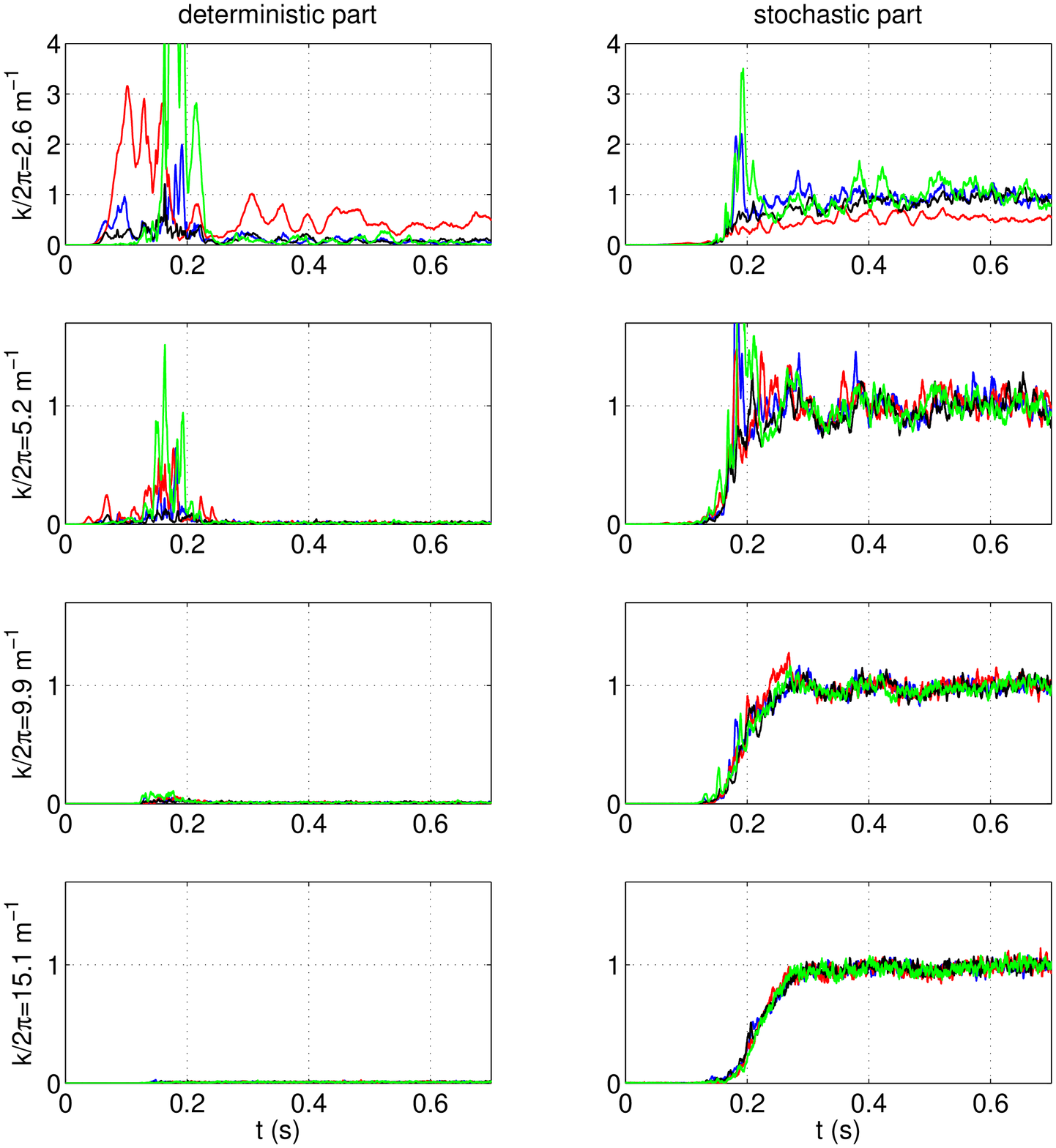}
\caption{Time evolution of the energy at various wave numbers for the 4 quadrants of propagation (blue: positive $x$ direction, red: positive $y$ direction, black: negative $x$ directions and green: negative $y$ direction). The spectrum has been shared in a deterministic part which is repeatable over realizations (left column) and a stochastic part (right column) (see text). The spectra have been normalized by the long time limit of the full spectrum.}
\label{spec3}
\end{figure*}

The forcing of the waves is a pure sine at a given frequency (30 Hz) which is fully coherent in time. Due to this deterministic and repeatable forcing scheme, the early steps of the propagation can be reproduced from one realization to another as well as the early nonlinear phase. After some time, the non linear process becomes chaotic and the inherent sensitivity to initial conditions leads to the development of a chaotic (or stochastic) stage that is distinct from one realization to another due to slight differences in the initial conditions. Furthermore, as observed above a part of the wave spectrum at all times is due to coherent radiation of waves by the shaker. We would then like to separate the coherent and repeatable part of the spectrum from the stochastic one as the latter part is the only one that may actually be described by the Weak Turbulence Theory. 

A way to distinguish the initial deterministic part of the motion from the subsequent stochastic one is the following. We compute the average $\langle v(\mathbf k,t)\rangle$. Non zero values of this average are possible only for the initial repeatable part of the development and for the part of the spectrum that remains coherent at the forcing frequency. The spectrum of the waves is then split into a deterministic (or repeatable) part 
\begin{equation}
E^{det}(\mathbf k,t)=|\langle v(\mathbf k,t) \rangle|^2
\end{equation}
and a stochastic part  
\begin{equation}
E^{sto}(\mathbf k,t)=\langle |v(\mathbf k,t)|^2 \rangle-|\langle v(\mathbf k,t) \rangle|^2 \,\, .
\end{equation}
This decomposition is somewhat reminiscent of the Reynolds decomposition of the velocity field in fluid turbulence into an average part and a fluctuating one.

Figure~\ref{spec3} shows the two parts of the spectrum. The deterministic part is seen to be significant at the lowest wavenumbers close to the forcing wavenumber ($k/2\pi\sim2.6$~m$^{-1}$). Furthermore the deterministic part is seen to increase strongly at the initial stage (i.e. when less than 250 ms elapsed since the shaker has been triggered) and then vanish except for the upward propagating wave that remains about 50\% of the total spectrum at this wave vector. The latter component displays some fluctuations that may be due either to a insufficient convergence of the statistics or to a dynamical interplay of the low frequency modes around $k/2\pi\sim2.6$~m$^{-1}$ that may interact non linearly and show a non constant dynamics. A weaker coherent part is also seen at $k/2\pi\sim5.2$~m$^{-1}$ possibly due to the resolution issue mentioned above.

The stochastic part of the spectrum starts to rise simultaneously for all directions. The duration of the rise is also similar for all directions except at the forcing wavenumber at which the stochastic dynamics is most likely affected by the deterministic one. The time at which the rise starts is increasing with the wavenumber.

\subsubsection{Summary}

The overall picture of the development of the spectrum is thus the following. In the initial stage, the forcing induces a deterministic linear wave propagating upwards. In a second step, the non linear transfers starts once the forcing wave hits the top boundary. A stochastic almost isotropic energy cascade is then initiated. The energy is transferred progressively to smaller and smaller scale. The rise of energy at the smallest scales is thus delayed compared to largest scales due to the time required for the nonlinear cascade to proceed. Once the stationary regime is attained a partially coherent upward propagating wave remains at the forcing frequency. There is almost no deterministic wave at the same frequency that propagates downwards. This suggests that the nonlinear energy exchanges with other waves destroy progressively the coherence of the forcing wave while it propagates across the plate. The coherence is most likely lost in a time comparable to the time needed to propagate along the plate.

\subsection{ A self similar non stationary solution ?}

Following Falkovich \& Shafarenko~\cite{Falkovich}, one looks for a front-like self-similar solution under the form: $n(k,t)=t^{-q}f(k/t^p)$. The scaling properties of the collision integral and the hypothesis that the stationary solution should be of constant flux leads to the solution 
\begin{equation}
n(\mathbf k,t)=\frac{1}{t}f\left(\frac{k}{\sqrt{t}}\right)\,\, ,
\label{sss}
\end{equation} 
where $f$ is an unknown function (see Ducceschi et al.~\cite{Ducceschi}). Thus one expects that a front will propagate at a cutoff wavenumber $k_c(t)\propto \sqrt{t}$. 

This has indeed been observed by Ducceschi et al. in a numerical simulation~\cite{Ducceschi}. By using a finite difference numerical scheme, they simulate a finite plate with realistic boundary conditions but without dissipation. They observe that the buildup of the turbulent spectrum occurs through a propagating front which cutoff frequency evolves as $\omega_c\propto t$ which, thanks to the dispersion relation $\omega\propto k^2$ translates into a cutoff wavenumber $k_c\propto \sqrt{t}$. As no dissipation is present in their simulation, the observed spectrum does not saturate to a stationary form. Humbert~\cite{Humbertphd} studied a simplified phenomenological model of the kinetic equation and observed the same behavior. Here we check to what extent this prediction is fulfilled in a real plate.

By discarding logarithmic corrections which are elusive in real systems, one expects the stationary solution to be $n^{KZ}(\mathbf k,t)\propto \frac{1}{k^2}$. One can rewrite the self-similar solution (\ref{sss}) into 
\begin{equation}
n(\mathbf k,t)=\frac{1}{k^2}g\left(\frac{k}{\sqrt{t}}\right)\,\, ,
\label{sss1}
\end{equation} 
where the function $g(\xi)$ is zero for $\xi\rightarrow\infty$ and goes to 1 as $\xi\rightarrow 0$. 

\begin{figure}[!htb]
\includegraphics[width=9.5cm]{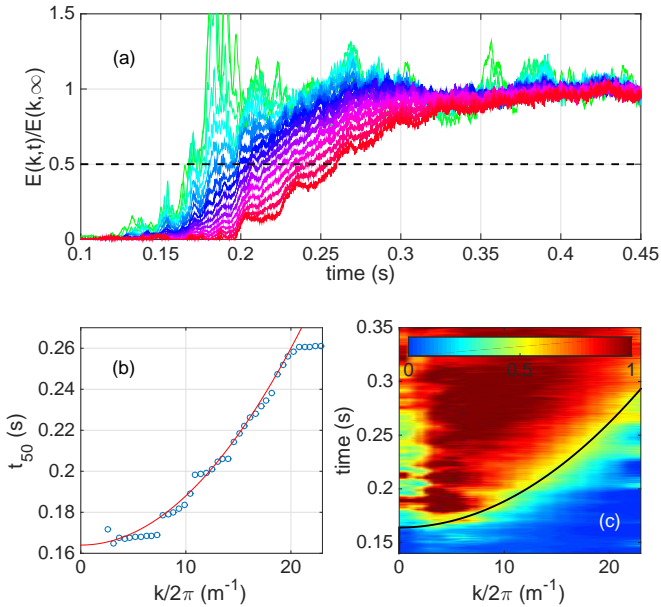}
\caption{Propagation of the energy front in the wavenumber space. (a) Evolution of the spectrum (averaged over directions) normalized by its final value $E(k,t)/E(k,+\infty)$ for a few selected values of $k$ equally spaced in the interval $[3.6,20]\times 2\pi$. Black dashed line: 50\% level defining the time $t_{50}(k)$ (symbols in (b)) required for the spectrum $E(k,t)$ (averaged over all wavevector directions) to reach half of its final energy. Red line in (b): linear fit of $t_{50}$ vs $k^2$. (c) $E(k,t)/E(k,+\infty)$. Black line: fit of $t_{50}$ as computed in (b).}
\label{front}
\end{figure}
In our experiments, we do not observe the Kolmogorov-Zakharov spectrum because of dissipation effects~\cite{R23,Humbert}. Instead of using equation (\ref{sss1}) we normalise the spectrum $E(k,t)$ (summed over the directions of $\mathbf k$) by its stationary asymptotic value. This normalized quantity $E(k,t)/E(k,\infty)$ evolves between 0 (initially) and 1 (at large times) so that it plays a role similar to the function $g$ in eq. (\ref{sss1}). Figure~\ref{front}(a,c) shows the normalized spectrum. It is quite clear that energy propagates from small to large $k$ with a front. To extract the position of the front, we compute at each value of $k$ the time $t_{50}(k)$ at which the spectrum reaches 50\% of its final value. This time is shown on figure~\ref{front}(b,c). $t_{50}(k)$ is seen to evolve quadratically with $k$ as predicted by the self-similar solution. The general expression of the self-similar solution (\ref{sss}) is due to the scaling properties of the interaction term $J_{\mathbf k,\mathbf k_1,\mathbf k_2,\mathbf k_3}$ and of the 4-wave interaction. Thus our observations support the kinetic equation approach (\ref{ke}) for the vibrating plate. It is also consistent with our previous observations of the stationary regime~\cite{Miquel3} which show that the wave turbulence regime observed in our plate is weakly nonlinear at moderate forcing.

\section{Numerical simulations}

The experiments presented above show some evidence that the transient phase is consistent with the WTT. Nevertheless there are a few difficulties in the experiment such as the locality of the forcing or the slight anisotropy of the spectrum or the fact that dissipation is weak but operating at all scales as reported in~\cite{R23,Humbert}. Another issue is that the measurement is made only on a fraction of the plate. We would like to complement the experiments by numerical simulations with similar amplitudes of forcing but in idealized conditions in order to test the robustness of our experimental observations. Our goal is not to perform a systematic study of the influence of dissipation on the development of the spectrum which is of great interest but beyond the scope of the present article. We focus on the case of a weak dissipation (comparable to experiments) that enables a non linear cascade to proceed.

\subsection{Numerical setup}

We perform numerical simulations of a plate with similar physical properties than those of the real plate. For details on the numerical code see Miquel et al.~\cite{Miquelphd,R23}. The motion is simulated using a pseudo-spectral code with periodic boundary conditions. The physical properties of the simulated plate are that of a $2\times2$~m$^2$ steel plate, 0.4~mm thick. The resolution is 192$^2$ grid points. Our goal is not to simulate accurately the real plate but rather to simulate the fundamental features that enable a qualitative comparison with the experiment. In that respect periodic boundary conditions are not realistic but they are much simpler to handle than the real boundary conditions and they allow us to use an efficient pseudo-spectral scheme.

Excitation is achieved by applying a force isotropically on the modes of wavenumber close to $k/2\pi=2.5$~m$^{-1}$. The force applied to each mode is oscillating at the linear frequency of the considerred mode so that to induce a secular growth of the amplitude in the linear stage. The phases of the forces of the excited modes are chosen randomly for each realization but kept constant in time for each single realization. This forcing scheme is similar to the experimental one in the fact that the forced modes are forced deterministically at a given constant frequency. A significant difference is that the forcing is homogeneous in space in the simulation whereas the force is applied locally in the experiment.

A linear dissipation is introduced in the simulated wave equation in which the dissipation rate is either (i) the dissipation rate measured experimentally~\cite{R24} (for a direct comparison with experiments) or (ii) a `localized' dissipation, for which the dissipation rate is vanishing in the interval $k\in[6,157]$~m$^{-1}$ ensuring a conservative cascade in this interval. Previous investigations have shown that such a `localized' dissipation (ii) allows us to observe a stationary spectrum in full agreement with the theoretical predictions of the Weak Turbulence Theory~\cite{R18,R23}. Indeed such dissipation is close to the canonical conditions assumed in the theory. Thus the prediction of a propagating conservative front should be verified when the front is in the conservative interval of wave numbers. When using the experimental dissipation (i), the computed turbulence is very close to the observed one (in the stationary regime) despite the very different boundary conditions~\cite{R23}. These previous observations show that the boundary conditions do not have a major impact.

Two configurations have been studied (see table~\ref{tab-param}) with both cases of dissipation. The forcing amplitudes have been chosen so that the root mean square velocity in simulations is similar to experimental values. The forcing scheme, the boundary conditions and the dissipation are distinct in each case so that a perfect match is not possible but the values of $z_{rms}$ and $v_{rms}$ are very close which is sufficient for a qualitative comparison with the experiment.

\begin{table}
\centering    
\begin{tabular}{c|c|c|c|c}
run & dissipation & $z_{rms} $ & $v_{rms} $ & $\#$ \\
 & & (mm)  & (m/s)  & realizations\\
\hline
A & exp.  & 2.2 & 0.5 & 300  \\
B & localized  &  1.5 & 0.43 & 588  \\
exp. & & 4.2 & 0.42 & 100\\
\end{tabular}
\caption{Parameters of the simulations. `exp' dissipation means that the dissipation rate is the same as the experimental one, `localized' dissipation means that an interval of wavenumbers exists with no dissipation (see text for details). For each run, the statistics have been averaged over a few hundred independent realizations of the forcing. $z_{rms} $ and $v_{rms} $ are the root mean square values of the deformation and the velocity in the stationary regime. In the experiment $z_{rms}=4.2$~mm and $v_{rms}=0.42$~m/s (very large scale deformation and motion in the experiment have been filtered out as they are not present in the simulation). Note that values of the velocity are very similar for run A, run B and the experiment.}
\label{tab-param}
\end{table}

In each case the forcing is started at $t=0$ with a plate which is perfectly flat. Then the code runs for 1~s of physical time. The numerical experiment is repeated several hundred times in order to ensure a good convergence of the statistics (see table~\ref{tab-param}). 

\subsection{Buildup of the numerical spectrum}

\subsubsection{Growth of the energy}

\begin{figure}[!htb]
\includegraphics[width=9.5cm]{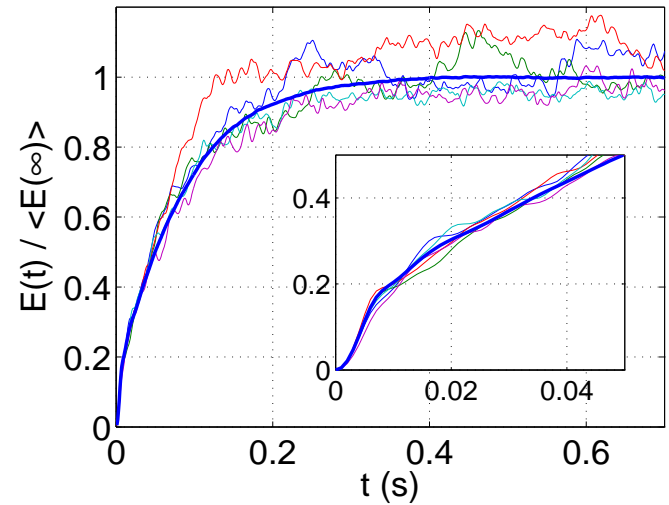}
\caption{Time evolution of the total energy for dataset A. Thick solid line: average over 300 realizations. Thin lines: 5 distinct realizations. Insert: zoom at short times. }
\label{Esim}
\end{figure}
The initial stage of the energy growth is somewhat distinct from that observed in the experiments due to the difference in the forcing scheme. In the simulations, the modes with $k/2\pi\approx 2.5$~m$^{-1}$ are forced resonantly at their linear eigenfrequency. Thus the initial stage is a fast linear secular growth of the amplitude of the forced modes. Accordingly the
energy grows initially quadratically with $t$ (fig.~\ref{Esim}). This growth is homogeneous in space so there is no initial propagation stage as is observed in the experiment making the initial transient stage easier to interpret. The first stage of the growth lasts until large enough amplitudes are reached so that the non linear terms become effective and initiate the energy cascade to small scales. The energy grows then almost linearly in time as expected for a constant input of energy. Because of the presence of dissipation, the spectrum finally saturates to a stationary value and the energy reaches a statistically stationary value: the average is constant but for individual realizations the energy fluctuates around the average value (fig.~\ref{Esim}). 

\subsubsection{Development of the spectrum}

\begin{figure}[!htb]
\includegraphics[width=9.5cm]{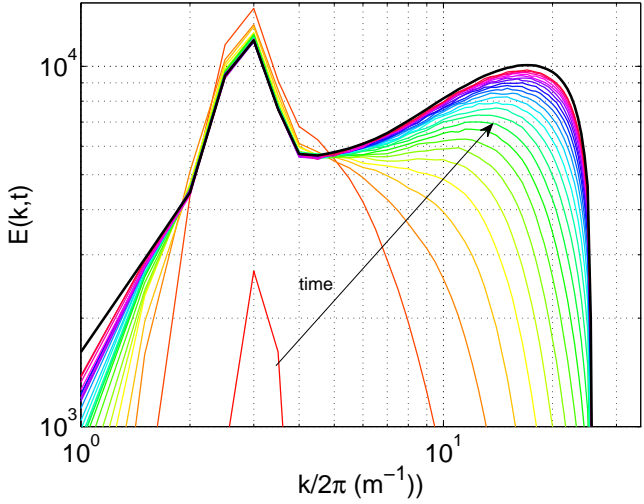}
\caption{Time evolution of the spectrum $E(k,t)$ for run A. The bottom most line corresponds to $t=4$ms and the subsequent lines are separated by a time interval $\delta t=10$~ms. }
\label{spsim}
\end{figure}
Figure~\ref{spsim} shows the time evolution of the angle integrated velocity spectrum $E(k,t)$ for run A. As expected from the evolution of the energy, the initial stage is simply a growth of the energy at $k/2\pi\approx 2.5$~m$^{-1}$. Then the turbulent state develops and energy propagates to high wave numbers. This behavior is very similar to that of the experiment (except for the initial propagation stage which is not present in the simulations). In the simulations the process is isotropic due to isotropic forcing. We expect the weak turbulence description to be valid only in the second part of the development of the spectrum, typically for $t>0.01$~s for the run A displayed in fig.~\ref{Esim} and \ref{spsim}.

\begin{figure}[!htb]
\includegraphics[width=9.5cm]{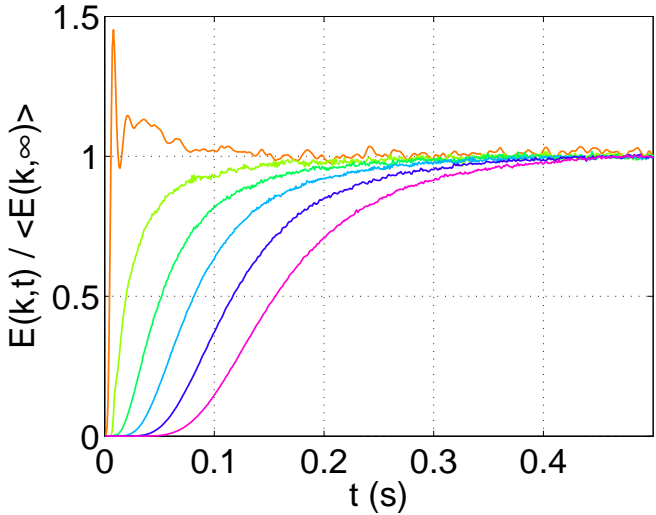}
\caption{Time evolution of the spectrum $E(k,t)/E(k,\infty)$ for various given values of $k$ for run A. The uppermost curve corresponds to $k/2\pi=2.5$~m$^{-1}$ which is the wavenumber of the energy input. Subsequent lines are separated by $\delta k/2\pi=5$~m$^{-1}$ with $k$ increasing from left to right.}
\label{spsim2}
\end{figure}
The time evolution of the normalized spectrum\\ $E(k,t)/ E(k,\infty)$ for a few values of $k$ is shown in fig.~\ref{spsim2}. The evolution of the energy of the modes that are directly excited by the forcing (left curve) is strongly different from what occurs at larger $k$. The energy of the forced modes grows rapidly due to the secular growth. It then overshoots the stationary value and displays a much slower decay to the limit value. This behavior is very similar to what is observed experimentally at the forced mode. 

The energy increase is delayed at larger $k$ compared to smaller values of $k$ in agreement with the idea of a front propagating in k-space. The much slower increase of the energy as compared to the secular growth of the forced mode is fully compatible with the phenomenology of weak turbulence. Indeed, the weakly non linear transfers are expected to occur on much slower time scales than the linear ones which is definitely what is observed here. This figure is actually very similar to figure~\ref{front}(a) for the experiment but the dynamics of the initial stage is simpler due to the isotropic forcing scheme. 

\subsection{Self similar non stationary solution ?}
\begin{figure}[!htb]
\includegraphics[width=9.5cm]{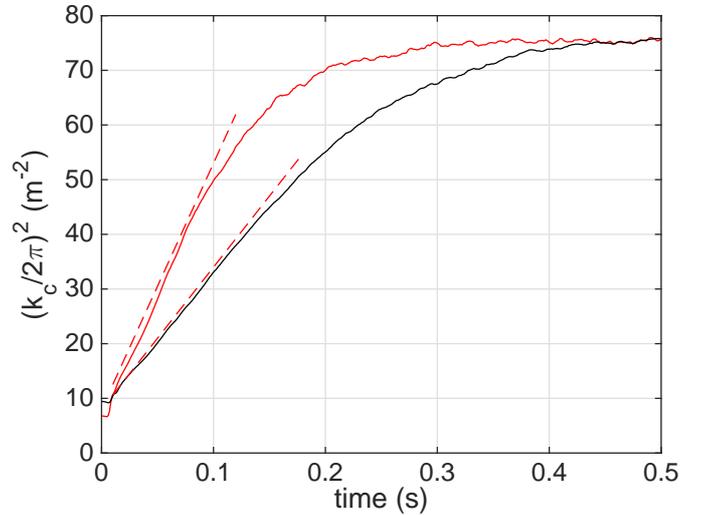}
\caption{Time evolution of $k_c$ for the numerical simulations. Red: experimental dissipation (run A). Black: localized dissipation (run B). Dashed lines are eye guides for a linear growth.}
\label{kcsimu}
\end{figure}

We would like to check if the simulation is compatible with the propagation of a self similar front as predicted by the theory.
Equation (\ref{sss1}) for the action spectrum translates for the angle integrated velocity spectrum as 
\begin{equation}
E(k,t)=kg\left(\frac{k}{\sqrt{t}}\right)\,\, .
\end{equation} 

Similarly to Ducceschi {et al.}~\cite{Ducceschi} we would like to define a cutoff wave number as
\begin{equation}
k_c(t)=\frac{\int kg(k/\sqrt{t})dk}{\int g(k/\sqrt{t})dk}\,\, ,
\end{equation}
which can be written as 
\begin{equation}
k_c(t)=\frac{\int E(k,t)dk}{\int E(k,t)/kdk}\,\, .
\label{kc}
\end{equation}
Unfortunately this expression can not be used with the experimental data because the very large peak at the forcing scale affects strongly the integrals in (\ref{kc}) and gives irrelevant values for $k_c$. The evolution of $k_c(t)^2$ for the numerical simulation is shown in fig.~\ref{kcsimu}. After a short transient (due to the initial linear dynamics), $k_c(t)^2$ increases almost linearly in both cases before saturating at a stationary value. For run B, we recover the result of Ducceschi  {\it et al.}~\cite{Ducceschi} until the front reaches the dissipative interval of wavenumber. For run A, we observe a linear evolution at intermediate times ($0.01\lesssim t\lesssim 0.1$s) which is both similar to that of run B and to the experimental observations.


\begin{figure}[!htb]
\includegraphics[width=9.5cm]{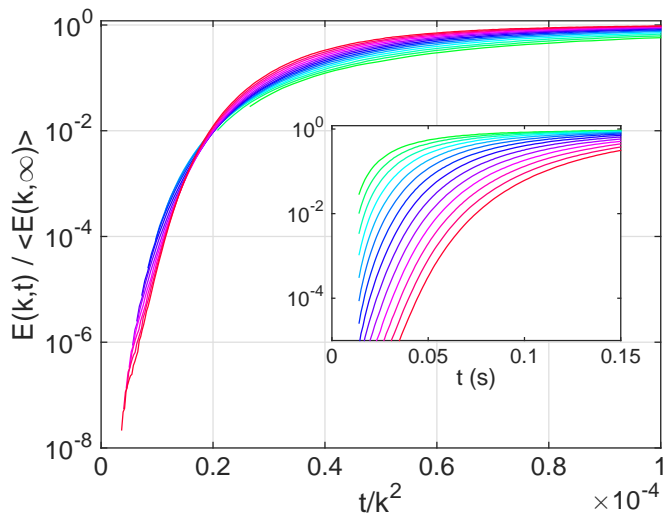}
\caption{Time evolution of the spectrum for the numerical simulations of run A as a function of the rescaled time $t/k^2$. In the insert the same curves are displayed as a function of time $t$. The color code is the same for both figures. The various curves correspond to wave numbers separated by $\delta k/2/\pi=1.5$~m$^{-1}$ the smallest value of $k$ is $k=14\pi$~m$^{-1}$. The evolution is plotted for $t>15 ms$ for times at which the energy cascades proceeds.}
\label{tcsimu}
\end{figure}

In fig.~\ref{tcsimu}, we show the increase of energy for run A for a collection of wave numbers in the inertial range. When time is rescaled by $k^2$, a fair collapse of the curves is observed at the initial stage of the growth (for small values of $t/k^2$). At larger times the final growth to the saturation does not rescale correctly. This shows a distinct dependency of the spectrum on the wavenumber. The initial growth is compatible with the prediction of the conservative cascade but the latest steps are influenced by dissipation and thus obey a different scaling (if any).

\section{Discussion and conclusion}

The weak turbulence theory is characterized by time scale separation. Indeed for weak non linearity the non linear time scale $T_{NL}$ is much larger than the linear period $T$ of the waves. In real systems, dissipation exists but if it remains weak then the dissipation time scale $T_d$ is much larger than $T_{NL}$. In this way, non linear energy exchange among waves can develop at intermediate time scales before dissipation begins to operate. Thus for $T\ll T_{NL} \ll T_d$ a regime of wave turbulence can develop. This time scale separation has been checked experimentally in the stationary regime by performing a wavelet decomposition~\cite{Miquel3}. 

In the transient regime, one does not expect that dissipation plays any major role until $t\sim T_d$. Due to the scale separation $T_{NL}\ll T_d$, a non stationary non linear regime has ample time to develop for $t< T_d$ which should be similar to the conservative case. Thus for times $t< T_d$ one expect to observe the conservative self similar solution. At later times $t\gtrsim T_d$, dissipation begins to operate and thus the propagation of the front should slow down and stop. The spectrum will slowly evolve to the stationary spectrum which depends on the dissipation in real plates.

For the real plate and for run A, the dissipation is active over the entire range of wavenumber but it is indeed weak. With the  experimental dissipation rate implemented in the simulation (measured in~\cite{R24}), the time scales associated with dissipation are of the order of $0.1$~s or larger. The experimentally measured non linear time scales are about one decade smaller than the dissipation time scales~\cite{Miquel3}. It means that a conservative non linear transient stage can indeed develop and will be faster than the dissipation that will have a significant impact on the vibration only after a time of order $0.1$~s. We then expect that the initial stage of the front propagation (for times less than $0.1$~s) should not be affected by the dissipation and should be consistent with the prediction of the WTT. This is indeed the case as the linear growth of $k_c^2$ is linear for times up to 0.1~s (fig.~\ref{kcsimu}). 
A qualitatively similar behavior is observed for run B. In this run, we implement a `localized' dissipation which is effective only at the highest wave numbers of the simulation. Thus the predicted self similar evolution of the spectrum is expected until the front in the transient spectrum reaches the wavenumber $k/2\pi=25$~m$^{-1}$. Above this wavenumber the dissipation becomes very large. Then the propagation of the front is affected by dissipation and the spectrum evolves toward the stationary solution as observed for $t>0.15$~s.

In conclusion, our experiments and numerical simulations show that the build up of turbulence follows a 3 steps sequence. 
\begin{enumerate}
\item First a very fast linear growth of the forced modes is observed due either to propagation of the forced wave from the shaker in the experiment or to secular growth of the forced modes in the simulation. This step is of course strongly dependent on the forcing scheme.
\item The second step corresponds to the start of the non linear energy cascade. The buildup of the turbulent spectrum occurs through the propagation of an almost isotropic front in the wave number space which is compatible with the self similar prediction based on the similarity properties of the kinetic equation (which has been predicted for the non dissipative case). This step seems more universal as soon as the time scale separation $T_{NL}\ll T_d$ is verified. 
\item The last step corresponds to the long time saturation of the spectrum. At times comparable to the dissipation time scale, dissipation comes into play and gradually stops the propagation of the front. The shape of the spectrum ultimately evolves slowly into its statistically stationary state. 
\end{enumerate}
The observed propagation of the front in the intermediate non linear stage is consistent with the scaling properties predicted by the Weak Turbulence Theory. Thus our observations support the general theoretical framework of weak turbulence leading to the kinetic equation for the time evolution of the wave spectrum. Such experimental support is extremely rare in the literature. In this respect the case of the vibrating plate is a precious model for weakly non linear wave turbulence in real systems and is a strong support to the relevance of the WTT for other real systems.

\thanks{MIA thanks the Argentinian government for financial support by the subsecreta\'ia de gesti\'on y empleo p\'ublico de la jefature de gabinete de ministros de la naci\'on. We thank J.-P. Thibault for sharing the anechoic room. The laboratory LEGI is part of the LabEx Tec 21 (Investissements d’Avenir - grant agreement n°ANR-11-LABX-0030).}

All authors contributed equally to the investigations reported in this article.

\bibliographystyle{epj}
\bibliography{plaquex}

\begin{thebibliography}{25}

\bibitem{R1}
V.E. Zakharov, V.S. L'vov, G.~Falkovich, \emph{Kolmogorov Spectra of
  Turbulence} (Springer, Berlin, 1992)

\bibitem{R2}
S.~Nazarenko, \emph{Wave Turbulence} (Springer, Berlin, 2011)

\bibitem{R3}
A.C. Newell, B.~Rumpf, Ann. Rev. Fluid Mech. \textbf{43} (2011)

\bibitem{Frisch}
U.~Frisch, \emph{Turbulence: the legacy of A.N. Kolmogorov} (Cambridge
  University Press, Cambridge, 1995)

\bibitem{R10}
S.~Nazarenko, S.~Lukaschuk, S.~McLelland, P.~Denissenko, J. Fluid Mech.
  \textbf{642}, 395 (2009)

\bibitem{R11}
E.~Falcon, C.~Laroche, S.~Fauve, Phys. Rev. Lett. \textbf{98}(9), 094503 (2007)

\bibitem{Sharon}
E.~Yarom, E.~Sharon, Nature Physics \textbf{10}, 510 (2014)

\bibitem{R8}
J.~Laurie, U.~Bortolozzo, S.~Nazarenko, S.~Residori, Physics Reports
  \textbf{514}, 121 (2012)

\bibitem{Boudaoud}
A.~Boudaoud, O.~Cadot, B.~Odille, C.~Touz\'e, Phys. Rev. Lett. \textbf{100},
  234504 (2008)

\bibitem{Mordant}
N.~Mordant, Phys. Rev. Lett. \textbf{100}, 234505 (2008)

\bibitem{R19}
P.~Cobelli, P.~Petitjeans, A.~Maurel, V.~Pagneux, N.~Mordant, Phys. Rev. Lett.
  \textbf{103}(20), 204301 (2009)

\bibitem{R21}
N.~Mordant, Eur. Phys. J. B \textbf{76}, 537 (2010)

\bibitem{R23}
B.~Miquel, A.~Alexakis, N.~Mordant, Phys. Rev. E \textbf{89}, 062925 (2014)

\bibitem{R24}
B.~Miquel, N.~Mordant, Phys. Rev. Lett. \textbf{107}(3), 034501 (2011)

\bibitem{R25}
B.~Miquel, A.~Alexakis, C.~Josserand, N.~Mordant, Phys. Rev. Lett.
  \textbf{111}, 054302 (2013)

\bibitem{Miquel3}
B.~Miquel, N.~Mordant, Phys. Rev. E \textbf{84}(6), 066607 (2011)

\bibitem{Humbert}
T.~Humbert, O.~Cadot, G.~D\"uring, C.~Josserand, S.~Rica, C.~Touz\'e, EPL
  \textbf{102}, 30002 (2013)

\bibitem{R18}
G.~D\"uring, C.~Josserand, S.~Rica, Phys. Rev. Lett. \textbf{97}, 025503 (2006)

\bibitem{Bedard}
R.~Bedard, S.~Lukaschuk, S.~Nazarenko, JETP Letters \textbf{97}(8), 459 (2013)

\bibitem{Falkovich}
G.E. Falkovich, A.V. Shafarenko, J. Nonlinear Sci. \textbf{1}, 457 (1991)

\bibitem{Ducceschi}
M.~Ducceschi, O.~Cadot, C.~Touz{\'e}, S.~Bilbao, Physica D-Nonlinear Phenomena
  \textbf{280-281}, 73 (2014)

\bibitem{R22}
A.~Maurel, P.~Cobelli, V.~Pagneux, P.~Petitjeans, Appl. Optics \textbf{48}, 380
  (2009)

\bibitem{Cobelli1}
P.J. Cobelli, A.~Maurel, V.~Pagneux, P.~Petitjeans, Exp. Fluids \textbf{46},
  1037 (2009)

\bibitem{Humbertphd}
T.~Humbert, Ph.D. thesis, Universit\'e Pierre et Marie Curie (2014)

\bibitem{Miquelphd}
B.~Miquel, Ph.D. thesis, Universit\'e Pierre et Marie Curie (2013)

\end{thebibliography}

 \end{document}